\documentclass{article}
\usepackage{spconf,amsmath,graphicx}

\title{CROSS-LINGUAL TRANSFER LEARNING FOR ALZHEIMER'S DETECTION FROM SPONTANEOUS SPEECH}
\name{Bastiaan Tamm, Rik Vandenberghe, Hugo Van hamme\sthanks{This research was supported by KU Leuven Special Research Fund grant C24M/22/025.}}
\address{KU Leuven, Belgium}
\begin{document}
%\ninept
%
\maketitle
\begin{abstract}
% 100-150 words
Alzheimer’s disease (AD) is a progressive neurodegenerative disease most often associated with memory deficits and cognitive decline. With the aging population, there has been much interest in automated methods for cognitive impairment detection. One approach that has attracted attention in recent years is AD detection through spontaneous speech. While the results are promising, it is not certain whether the learned speech features can be generalized across languages. To fill this gap, the ADReSS-M challenge was organized. This paper presents our submission to this ICASSP-2023 Signal Processing Grand Challenge (SPGC). The model was trained on 228 English samples of a picture description task and was transferred to Greek using only 8 samples. We obtained an accuracy of 82.6\% for AD detection, a root-mean-square error of 4.345 for cognitive score prediction, and ranked 2nd place in the competition out of 24 competitors.

\end{abstract}
\section{Introduction}
\label{sec:intro}

Alzheimer’s disease (AD) is a progressive neurodegenerative disease most often associated with memory deficits and cognitive decline. It is the most common form of dementia and the fifth-leading cause of death among people age 65 or older~\cite{alz2022}.
With an aging population, there has been much interest in automated methods for cognitive impairment detection, especially ones that are inexpensive and easily scalable. One possibility that has gained a lot of attention in recent years is to analyze spontaneous speech, a readily available medium that can provide insight into the working of the brain. However, most of the proposed approaches have not investigated which speech features can be transferred across languages for AD detection~\cite{delaFuenteGarcia2020}.

 Hence, ADReSS-M, an ICASSP-2023 Signal Processing Grand Challenge (SPGC), was organized to investigate this matter~\cite{luz23}.
The goal is to train a model on English speech from a picture description task and apply it to a different picture description task in Greek. The challenge has two tasks, first to predict the AD diagnosis and second to predict the Mini-Mental State Examination (MMSE) score of a participant.
%a 30-point questionnaire used extensively in clinical and research settings to measure cognitive impairment.
The latter is a questionnaire that is used extensively in clinical and research settings to measure cognitive impairment and is scored out of 30 points.

In this paper, we present our submission to the challenge.\footnote{Our Code: https://github.com/lcn-kul/madress-2023} The models use a sequence of acoustic features and covariates (age, gender, education) to make the predictions. They are first trained in English, and then they are transferred to Greek using mixed-language batches and parameter averaging. This approach obtained an accuracy of 82.6\% for AD detection and 4.345 for cognitive score prediction, compared to 73.9\% and 4.955 respectively by the baseline. With this submission, we ranked 2nd place in the competition out of 24 competitors.

\section{Methods}
\label{sec:methods}

\subsection{Datasets}
\label{ssec:datasets}

Two datasets were used in this challenge, one English and one Greek, consisting of audio recordings of healthy controls and AD patients who were asked to describe a picture. The challenge organizers divided the data into three splits: an English training split (n=237, 122~AD), a Greek sample split (n=8, 4~AD), and a Greek test split (n=46, 22~AD). The test statistics were derived from the confusion matrix of the submission and were not known ahead of time. It was known that the splits were balanced for AD, age, and gender~\cite{luz23}.

We removed one healthy control from the English training split (no cognitive score) and 8 AD patients for balancing (n=228, 114~AD). Finally, in 12 controls where education was not available, we assumed the missing value to be 12 years.

\subsection{Preprocessing}

Each audio file is split into ten equal segments.
% https://sail.usc.edu/publications/files/eyben-preprinttaffc-2015.pdf
% https://doi.org/10.1145/1873951.1874246
For each segment, a 25-D eGeMAPS~\cite{eyben2015geneva} feature vector is calculated using openSMILE~\cite{opensmile}.
%A 25-D eGeMAPS feature vector is calculated for each segment using openSMILE.
%The feature level is set to ``low-level descriptors" to keep the input dimension small. This results in a sequence of ten 25-dimensional vectors for each audio file.

% Each audio file is split into ten equal segments. U
% % https://sail.usc.edu/publications/files/eyben-preprinttaffc-2015.pdf
% % https://doi.org/10.1145/1873951.1874246
% For each segment, a 25-D eGeMAPS~\cite{eyben2015geneva} feature vector is calculated using% openSMILE~\cite{opensmile}.
% %A 25-D eGeMAPS feature vector is calculated for each segment using openSMILE.
% %The feature level is set to ``low-level descriptors" to keep the input dimension small. This results in a sequence of ten 25-dimensional vectors for each audio file.

\subsection{Model}
\label{ssec:model}

Both the AD detection model and the cognitive score prediction model are based on the same architecture.
They are rather small: only 767 and 468 parameters respectively.
Each model takes as input a sequence of ten eGeMAPS features and the covariates age, gender and education. The estimated AD probability is also included as a covariate for the cognitive score prediction model. For simplicity, the covariates are concatenated to the eGeMAPS feature sequence.

The architecture consists of four parts. First, batch normalization is applied to the input features. Next, the features are down-projected into a smaller hidden space (12- or 8-dimensional respectively), followed by dropout and ReLU activation.

Afterward, attention pooling is used to collapse the time dimension. The attention weights are calculated using a 2-layer feed-forward network with an intermediate space that is twice as large as the hidden space, followed by softmax such that the weights sum to 1. Between the two layers, dropout and ReLU activation are used.

The final step is a linear projection to map the vector to the output space (2- or 1-dimensional respectively). For the cognitive score prediction model, a sigmoid function is used to map the value to the range [0,1]. The MMSE labels are also normalized to the same range.

\subsection{English pre-training}
\label{ssec:pretraining}

%The training procedure is identical for both the AD detection and the cognitive score prediction models.

The English data is split into 80\% train and 20\% validation.
%(used in the next section).
The models are trained on the English training data and validated on the Greek sample set. To account for the effects of random initialization, training takes place five times with five different random seeds. The model with the lowest validation loss over the five runs is selected as the pre-trained model.

%The English validation data is not used in this stage, instead the models are validated on the Greek sample set. The English validation set will be utilized in the following section.

%Training takes place five times with five different random seeds. The model with the lowest validation loss over the five runs is selected as the pre-trained model. 

\subsection{Mixed-batch transfer learning}
\label{ssec:transfer_learning}

The pre-trained model is finetuned on the English training data and 4 of the Greek samples (2 AD). Every fifth sample of the mini-batch is replaced by a Greek sample. 
% the other 4 are held out for validation. The training continues with the English training data, but now Greek samples are inserted into the mini-batches every fifth sample.
The model is validated on the English validation data and the 4 held-out Greek samples, inserted in the same way.
%The model with the lowest validation loss is selected.

To improve robustness, we repeat the procedure but swap the 4 training Greek samples with the 4 held-out samples. Finally, the parameters of these two models can be averaged since they are initialized from the same pre-trained model.

\subsection{Training details}
\label{ssec:training_details}

The models are implemented using the PyTorch (v.1.11.0) and PyTorch Lightning (v.1.8.6) libraries in Python 3.8.
The network is trained using the AdamW optimizer
%~\cite{loshchilov2017decoupled}
with a weight decay of 1e-2. The learning rate is warmed up linearly for 100 steps and is fixed at 3e-3 afterward. Cross-entropy loss is used for the AD detection model, and mean-square-error loss is used for the cognitive score prediction model. Each model is trained with a batch size of 32 for a total of 30 epochs, and the model with the lowest validation loss is selected.

\section{Results}
\label{sec:results}

The challenge allowed for five submissions. So, to test the robustness of the procedure proposed above, the entire procedure was run five times with different random seeds. The test accuracies for the AD detection task in ascending order were 71.7\%, 73.9\%, 76.1\%, 80.4\% and \textbf{82.6\%}. The best-performing model had a specificity of 91.7\%, precision of 88.9\%, sensitivity of 72.7\%, and an F1-score of 80.0\%.

The root-mean-square errors (RMSE) of the cognitive score prediction task in descending order were 4.837, 4.816, 4.716, 4.713, \textbf{4.345}. Note that these models use the probabilities of the AD detection model as input. Since the best AD detection model was not known ahead of time, the probabilities of the 5 submitted AD detection models were averaged.

\section{Conclusions}

In this paper, we present our submission for the \mbox{ADReSS-M} challenge. Our approach outperforms the best baseline model and is robust to model initialization. The innovation in our work lies in pre-training and mixed-batch fine-tuning procedure. The actual architecture is extremely simple and we believe with an improved feature extraction network, performance can be further improved. 

\bibliographystyle{IEEEbib}
\bibliography{strings,refs}

\end{document}